\documentclass[runningheads]{llncs}

\usepackage{graphicx}
\usepackage{hyperref}

\usepackage[misc]{ifsym}

\usepackage{microtype}
\usepackage{wrapfig}
\usepackage{cite}

\usepackage{subcaption}

\usepackage{listings}
\usepackage{color}

\usepackage{xcolor}
\hypersetup{
	colorlinks,
	linkcolor={red!50!black},
	citecolor={blue!50!black},
	urlcolor={blue!80!black}
}

\definecolor{codegreen}{rgb}{0,0.6,0}
\definecolor{codegray}{rgb}{0.5,0.5,0.5}
\definecolor{codepurple}{rgb}{0.58,0,0.82}
\definecolor{backcolour}{rgb}{0.95,0.95,0.92}
 
\lstdefinestyle{mystyle}{
    commentstyle=\color{codegreen},
    keywordstyle=\color{magenta},
    numberstyle=\tiny\color{codegray},
    stringstyle=\color{codepurple},
    basicstyle=\scriptsize, 
    breakatwhitespace=false,         
    breaklines=true,                 
    captionpos=b,                    
    keepspaces=true,                 
    numbers=left,                    
    numbersep=5pt,                  
    showspaces=false,                
    showstringspaces=false,
    showtabs=false,                  
    tabsize=2
}
 
\usepackage{amsmath}
\usepackage{amssymb}

\begin{document}
\title{Functional Federated Learning\\ in Erlang (\texttt{ffl-erl})\thanks{The final authenticated version is available online at \url{https://doi.org/10.1007/978-3-030-16202-3_10}.}}
\titlerunning{Functional Federated Learning in Erlang}
%

\author{Gregor Ulm$^{(\textrm{\Letter})}$ \inst{1, 2}
\orcidID{0000-0001-7848-4883}
\and
Emil Gustavsson \inst{1, 2}
\orcidID{0000-0002-1290-9989}
\and
Mats Jirstrand \inst{1, 2}
\orcidID{0000-0002-6612-8037}
}
\authorrunning{G. Ulm et al.}

\institute{Fraunhofer-Chalmers Research Centre for Industrial Mathematics,\\ Chalmers Science
Park, 412 88 Gothenburg, Sweden\\
\and Fraunhofer Center for Machine Learning,
\\Chalmers Science
Park, 412 88 Gothenburg, Sweden\\
\email{\{gregor.ulm,emil.gustavsson, mats.jirstrand\}@fcc.chalmers.se}
\url{http://www.fcc.chalmers.se/}
}

\maketitle              
\begin{abstract}
The functional programming language Erlang is well-suited for concurrent and distributed applications, but numerical computing is not seen as one of its strengths. Yet, the recent introduction of Federated Learning, which leverages client devices for decentralized machine learning tasks, while a central server updates and distributes a global model, motivated us to explore how well Erlang is suited to that problem. We present the Federated Learning framework \texttt{ffl-erl} and evaluate it in two scenarios: one in which the entire system has been written in Erlang, and another in which Erlang is relegated to coordinating client processes that rely on performing numerical computations in the programming language C. There is a concurrent as well as a distributed implementation of each case. We show that Erlang incurs a performance penalty, but for certain use cases this may not be detrimental, considering the trade-off between speed of development (Erlang) versus performance (C). Thus, Erlang may be a viable alternative to C for some practical machine learning tasks.

\end{abstract}

\section{Introduction}
With the explosion of the amount of data gathered by networked devices, more efficient approaches to distributed data processing are needed. The reason is that it would be infeasible to transfer all data gathered from edge devices to a central data center, process it, and afterwards transfer results back to edge devices via the network. There are several approaches to taming the amount of data received, such as filtering on edge devices, transferring only a representative sample, or performing data processing tasks decentrally. A recently introduced example of distributed data analytics is Federated Learning~\cite{mcmahan2016communication}. Its key idea is the distribution of machine learning tasks to a subset of available devices, followed by performing machine learning tasks locally on data that is available on edge devices, and iteratively updating a global model.

In this paper, we present \texttt{ffl-erl}, a Federated Learning framework implemented in the functional programming language Erlang.\footnote{Source code artifacts accompanying this paper are available at \url{https://gitlab.com/fraunhofer_chalmers_centre/functional_federated_learning}.} This work was produced in the context of an industrial research project with the goal of exploring and evaluating various approaches to distributed data analytics in the automotive domain. Our contribution consists of the following:

\begin{itemize}
\item Creating \texttt{ffl-erl}, the first open-source implementation of a framework for Federated Learning in Erlang
\item Highlighting the feasibility of functional programming for the aforementioned framework
\item Creating a purely functional implementation of an artificial neural network in Erlang
\item Comparing the performance of a Federated Learning implementation fully in Erlang with one in which client processes are implemented in C
\item Exploring two approaches of integrating C with Erlang: NIFs and C nodes
\end{itemize}

The remainder of our paper is organized as follows: Section~\ref{problem} contains background information and describes the motivating use case. Section~\ref{solution} covers our implementation in detail and presents experimental results. Section~\ref{related} gives a brief overview of related work, while Section~\ref{future} describes future work. Appendix~\ref{appendix_A} contains a mathematical derivation of Federated Stochastic Gradient Descent.

\section{Background}
\label{problem}
This section gives an overview of Federated Learning (\ref{prob_1}) and presents the mathematical foundation of Federated Stochastic Gradient Descent (\ref{prob_2}). It furthermore describes our motivating use case (\ref{prob_3}).

\subsection{Federated Learning}
\label{prob_1}
Federated Learning is a decentralized approach to machine learning. The general idea is to perform machine learning tasks on a potentially very large number of edge devices, which process data that is only accessible locally. A central server is relegated to assigning tasks and updating the global model based on the local models it receives from edge devices. One iteration of Federated Learning consists of the following steps, following McMahan et al.~\cite{mcmahan2016communication}:

\begin{enumerate}
\item Select a subset $c$ of the set of clients $C$
\item Send the current model from the server to each client $x\in c$
\item For each $x$, update the provided model based on local data by performing iterations of a machine learning algorithm 
\item For each $x$, send the updated model to the server
\item Aggregate all received local models and construct a new global model 
\end{enumerate}

There are several motivations behind Federated Learning. First, there is the bandwidth problem in a big data setting. The amount of data generated by local devices is too large to be transferred via the network to a central server for processing. Second, edge devices are getting more and more powerful. Modern smartphones, for instance, have been compared to (old-generation) supercomputers in our pockets in terms of raw computational power~\cite{bauer2012supercomputer}. Therefore, it seems prudent to more efficiently use these resources. Third, there are data privacy issues, as some jurisdictions have strict privacy laws. Thus, transmitting data via the network in order to perform central machine learning tasks is frayed with data privacy issues. This is summarized by Chen et al.~\cite{chen2012data}, while Tene et al.\ point out legal issues~\cite{tene2011privacy}. Federated Learning sidesteps potential legal quagmires surrounding data privacy laws and regulations as data is not centrally collected.

\subsection{Federated Stochastic Gradient Descent}
\label{prob_2}
Federated Stochastic Gradient Descent (Federated SGD) is based on Stochastic Gradient Descent (SGD), which is a well-established method in the field of statistical optimization. We first describe SGD, followed by a presentation of Federated SGD. The latter is based on McMahan et al.~\cite{mcmahan2016communication}.

\subsubsection{Stochastic Gradient Descent.}
\label{sub:sgd}

The aim of SGD is to minimize an objective function $F$ that is defined as the following sum:

\begin{equation}
F(w) = \frac{1}{n} \displaystyle\sum_{i=1}^{n} F_i(w).
\end{equation}

The goal is to find a value for the parameter vector $w$ that minimizes $F$. The value $F_i$ represents the contribution of element $i$ of the input data to the objective function. In order to minimize $F$, the gradient $\nabla$ is computed. The learning rate~$\eta$ is a factor that adjusts how far along the gradient the parameter update step is taken. It modifies the magnitude of change of $w$ between iterations. The parameter $w$ is updated in the following way:

\begin{equation} \label{eq:2}
w := w - \eta \nabla F(w).
\end{equation}

This means that the parameter $w$ is updated by computing the gradient of the objective function, evaluated for the previous parameter value, which is subtracted from the previous parameter value $w$. Since $F$ is a separable function, Eq.~\ref{eq:2} can be reformulated as

\begin{equation} \label{eq:3}
w := w - \frac{\eta}{n} \displaystyle\sum_{i=1}^{n} \nabla F_i(w).
\end{equation}

As indicated before, the learning rate $\eta$ is a modifier for slowing down or speeding up the training process. In practice, small positive values in the half-closed interval $(0, 1]$ are used. A common starting value is $0.01$. A learning rate that is too high may overshoot a global optimum. A learning rate that is too low, on the other hand, may severely impact the performance of the algorithm.

\subsubsection{Federated Stochastic Gradient Descent.}
\label{sub:fsgd}
Federated SGD is an extension of SGD. It takes into account that there are $k$ partitions $P_j$ of the training data, with $j$ ranging from $1$ to $k$, i.e.\ there is a bijection between partitions and clients. Consequently, Eq.~\ref{eq:3} has to be modified as we need to consider the work performed on each client $\in c$, where $c$ is the chosen subset of all clients $C$. The objective function is attempted to be minimized for each of the $k$ clients. However, the goal is to optimize the global model, not any of the local models. For Eq.~\ref{eq:4}, keep in mind that there are $n$ elements in the input data, thus $n = \sum\nolimits_{j}|P_j|$.

\begin{equation} \label{eq:4}
F^{j}(w) = \frac{1}{|P_j|} \displaystyle\sum_{i \in P_{j}}^{} F_i(w) \mathit{~for~} j = 1,\dotsc,k
\end{equation} The global objective function is shown in Eq.~\ref{eq:5}. Its full derivation is provided in Appendix~\ref{appendix_A}.
\begin{equation} \label{eq:5}
F(w) = \frac{1}{n} \displaystyle\sum_{j=1}^{k} |P_j| F^{j}(w)
\end{equation}

\subsection{Motivating Use Case}
\label{prob_3}
Intelligent vehicles generate vast amounts of data. According to recent industry figures, they can generate dozens of gigabytes of data per hour~\cite{coppola2016connected}. Considering even a moderately sized fleet of just a few hundred cars, collecting data, transferring data to a central server, processing data on a central server and afterwards sending results to each car is infeasible as we are already in the region of terabytes of data per hour. Yet, even simple tasks like filtering on the client can provide valuable insights. This is an example of a relatively straightforward way of reducing input data to a small fraction of its original volume, which highlights the importance of decentralized data processing.

However, our focus is on a more complex use case in the context of distributed data analytics. We explore training a machine learning model on client devices with local data, while a central server performs supplementary tasks. This relates to a real-world setting in which connected cars~\cite{evans2005connected} are equipped with on-board units that continuously gather data. These on-board units are general-purpose computers with performance metrics comparable to smartphones. For instance, our hardware uses an ARM-based multi-core CPU, similar to those found in a typical mid-range smartphone. On-board units are connected via wireless or 4G broadband networking to a central server, possibly via intermediaries, so-called road-side units. This is by no means a merely theoretical scenario. For instance, a recent large-scale experiment with road-side units was carried out by Lee and Kim~\cite{lee2010roadside} in South Korea in 2010.

\section{Solution}
\label{solution}
Our research prototype simulates a distributed system in which a central server interacts with a large number of clients. We first describe the main components of the framework itself~(\ref{Sol1}). This is followed by a discussion of a purely functional implementation of an artificial neural network (ANN) in Erlang~(\ref{Sol2}). Subsequently, we describe how the skeleton and the ANN can be combined~(\ref{Sol3}). Finally, we discuss experimental results~(\ref{Sol4}).

\subsection{The Skeleton of the Framework}
\label{Sol1}
This section illustrates the main ideas behind implementing a distributed machine learning framework. Consequently, we present the main parts of our skeleton, i.e.\ the client and server processes. The source code in this section leaves some details unspecified, but these can be filled in easily or referenced in the accompanying code repository. It seems appropriate to preface the discussion of our source code by briefly explaining the communication model of Erlang. In Erlang, processes communicate asynchronously by sending messages to each other. Each process has its own mailbox for incoming messages, which are processed in the order they arrive. However, the order in which they arrive is non-deterministic. If process $C$ receives one message each from processes $A$ and $B$, in this order, there is no guarantee that they were also sent in this order.

The skeleton consists of a client process, which may be instantiated an arbitrary number of times, and a server process. Both are shown in Code Listing~\ref{skeleton_client}. In the client process, the \texttt{receive} clause awaits a tuple tagged with the atom \texttt{assignment}. The received model is trained with local data via the function \texttt{train}. Examples of such a model are the weights of an ANN or parameters of a linear regression equation. After training has concluded, the updated model is sent to the server process. A tuple that is tagged with the atom \texttt{update} is sent to the server, using the operator '\texttt{!}', which is pronounced as \emph{send}. The server is addressed via the process identifier \texttt{Server\_Pid}. Thus, line 5 has to be read from right to left to trace the execution, i.e.\ we take a tuple tagged as \texttt{update}, containing the process identifier of the current process that is returned when calling the function \texttt{self} as well as the new local model, and send it to the server identified by \texttt{Server\_Pid}. Afterwards, the client function is called recursively, awaiting an updated model.
  
\begin{figure}
\begin{lstlisting}[language=Erlang, caption=Client and Server processes, label=skeleton_client]
client() ->
  receive
  { assignment, Model, Server_Pid } ->
    Val = train(Model), % computes 'w_j'
    Server_Pid ! { update, self(), Val },
    client()
  end.

server(Client_Pids, Model) ->
  Subset = select_subset(Client_Pids),
  % Send assignment: 
  [ X ! { assignment, Model, self() } || X <- Subset ],
  % Receive values:
  Vals = [ receive { update, Pid, Val } -> Val end || Pid <- Subset ],
  % Update model, i.e. compute global 'w':
  Model_ = update_model(Model, Vals, length(Client_Pids)),
  % Note: it is a simplification to use the number of clients; in this
  % case, each client has the same number of data points to work with
  server(Client_Pids, Model_).  
\end{lstlisting}
\end{figure}

The server process shown in Code Listing~\ref{skeleton_client} does not perform computationally intensive tasks. Instead, its role is to maintain a global model, based on updates received from client processes. Our system selects a random subset of all available devices. Sending the current model to the selected subset of client processes can be concisely expressed via mapping over a list or a list comprehension. It is assumed that all devices complete their assignments. This is reflected in the list comprehension in line 14, which blocks until the results of all assignments have been received. The resulting list of values \texttt{Vals} contains the updated local models of all client processes, with which a new global model will be constructed. The corresponding function \texttt{update\_model} is unspecified, however. After updating the model, the server process calls itself recursively. Overall, the preceding code is a textbook case of message passing in Erlang.

\subsection{A Neural Network with Backpropagation in Erlang}
\label{Sol2}
\subsubsection{Artificial Neural Networks.}

Artificial Neural Networks (ANNs) are a standard method in machine learning for a variety of learning tasks. A prime example is classification based on pattern recognition, for instance tagging images with keywords. The general principle is to minimize an objective function that computes the magnitude of an error. There are normally three steps to deploying an ANN: training, validation, and use in production. First, a labeled data set is used to train an ANN, which has the goal of minimizing the objective function. There is the risk that the ANN has been over-trained, i.e.\ it has memorized its input. Therefore, a labeled validation set is used to ensure that a data set that is similar to the test set is also correctly classified. If those two steps have been performed satisfactorily, the ANN is ready to be used for real-world data classification tasks.

\begin{wrapfigure}[8]{R}{6.5cm}
\centering
  \includegraphics[scale=0.30, trim={2cm 70 1 2cm},clip]{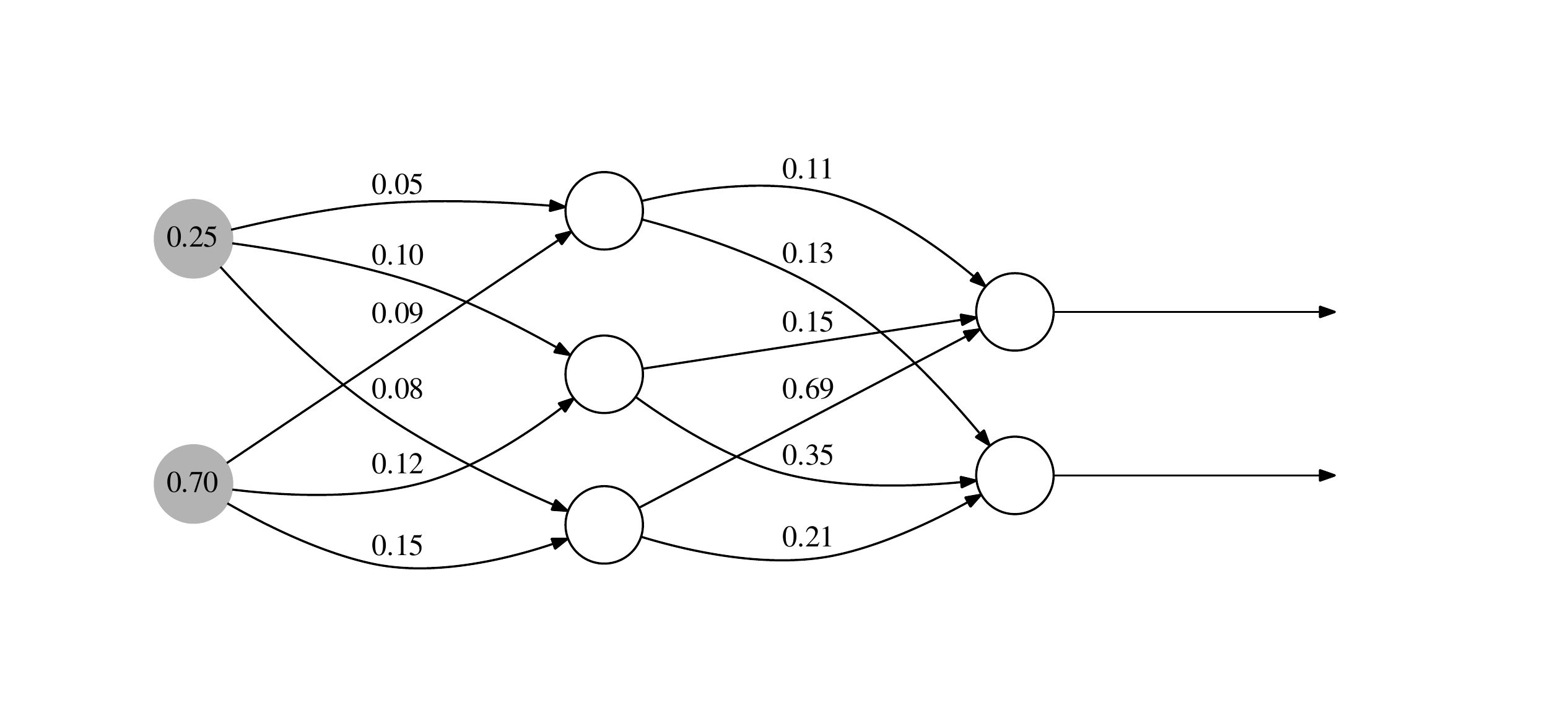}
  \caption{Artificial neural network}
  \label{fig:ann}
\end{wrapfigure}

Figure~\ref{fig:ann} shows a typical ANN. It consists of two input neurons, three hidden neurons, and two output neurons. The two input neurons on the left are shaded in order to indicate that an ANN is normally not applied to fixed input values but instead applied sequentially to each element of a larger data set. The layer of neurons in the middle is the hidden layer; the layer on the right is the output layer. The edges labeled with their weights represent connections between neurons. The edges leaving the output layer transmit the final output. There are two sets of labeled edges, one set connecting the input layer to the hidden layer and the other connecting the hidden layer to the output layer. Edge weights are initialized to a small random value and updated via training. The goal is to minimize the output error, which is based on the difference between the target values and the values the output layer neurons emit. After a forward pass we can determine how close the values emitted by the output neurons are to the target values. This is followed by adjusting the weights of the ANN with the backpropagation algorithm. Together, these two steps amount to one \emph{epoch}. In the end, the output error is minimized via iterative adjustments of the weights of the ANN.

Using the ANN in Fig.~\ref{fig:ann} as an example, we first perform a forward pass, which consists of computing the input of each hidden layer neuron by calculating the dot product of the input weights and all edges connecting the input nodes with that hidden layer neuron. For instance, the input of the topmost hidden layer neuron is $0.25 \times 0.05 + 0.70 \times 0.09 = 0.0755$. After applying the \emph{activation function} to that value, the input and output values of the output layer neurons are computed similarly. The activation function computes the output of a node, taking its input as the argument. Afterwards, the difference between target and actual output values can be calculated. This is followed by a backpropagation pass, in which the weights of the ANN are updated: first the weights of the edges from the output layer to the hidden layer, then the weights of the edges from the hidden layer to the input layer. These calculations are similar to the forward pass, except that the gradient, i.e.\ the derivative of the objective function we want to minimize, is used when calculating the respective dot products. Training an ANN with a batch of input data is done by processing all elements of the provided data, using them one by one as input for the input layer and performing one epoch. After each iteration, the weights are retained as the goal is to train on the entire set of input data.

For the sake of brevity, our description of an ANN does not consider common modifications such as setting a specific learning rate or using adaptive behavior based on previous results. We furthermore use a standard activation function, the \emph{sigmoid function}. Practitioners may use different activation functions or resort to various engineering techniques for improving the performance of ANNs as described, for instance, by Orr et al.~\cite{orr2003neural}. As a final note, we would like to highlight that ANNs can approximate any function~\cite{gybenko1989approximation, hornik1991approximation}, which is commonly referred to as the universal approximation theorem. Consequently, ANNs are widely used in practice. The example described above, consisting of three layers, is a shallow ANN. Those are versatile, but they are not efficient for large and very complex problems. A particularly noteworthy early breakthrough of shallow ANNs was the successful classification of handwritten digits, which is used by postal services~\cite{srihari1997integration}. More recent developments include deep neural networks, often referred to as deep learning. Those are ANNs with multiple hidden neuron layers, consisting of large numbers of neurons. 

\subsubsection{Implementation.}
In the following, we cover some aspects of an exemplary implementation of a basic ANN in Erlang. We will again leave out some implementation details, and instead focus on the big picture.\footnote{For illustrative purposes, we chose clear code over computationally more efficient code at some points. For instance, the function \texttt{forward} in Code Listing~\ref{annHelper} constructs a temporary list, which could be avoided by computing the dot product with an accumulator. However, for benchmarking purposes we used more efficient code.} Code Listing~\ref{ann_core} shows the function \texttt{ann}, which models an artificial neural network. The various helper functions it calls are shown in Code Listing~\ref{annHelper}. The input of the function \texttt{ann} consists of the values of the input neurons \texttt{Input}, the weights of both layers \texttt{Weights}, and the target values of the output layer~\texttt{Targets}.

\begin{figure}
\begin{lstlisting}[language=Erlang, caption=The core ANN function, label=ann_core]
ann(Input, Weights, Targets) ->
  { W_Input, W_Hidden } = Weights,
  % Forward Pass:
  Hidden_In  = forward(Input, W_Input, []),
  Hidden_Out = [ activation_fun(X) || X <- Hidden_In ],
  Output_In  = forward(Hidden_Out, W_Hidden, []),
  Output_Out = [ activation_fun(X) || X <- Output_In ],
  % Target vs. output:
  Delta = lists:zipwith(fun(X, Y) -> X - Y end, Targets, Output_Out),
  % Reverse pass:
  Output_Errors = output_error(Output_Out, Targets),
  % Update weights for output layer:
  W_Hidden_  = backpropagate(Hidden_Out, Output_Errors, W_Hidden,  []),          
  Hidden_Err = errors_hidden(Hidden_Out, Output_Errors, W_Hidden_, []),
  W_Input_   = backpropagate(Input, Hidden_Err, W_Input, []),
  { Output_Errors, { Input, { W_Input_, W_Hidden_ }, Targets } }.
\end{lstlisting}
\end{figure}

As described earlier, as a first step the ANN computes the input of the hidden layer. The output of the hidden layer is the result of mapping the activation function over the list \texttt{Hidden\_In}; the corresponding values of the output layer are computed in the exact same way. The list \texttt{Delta} contains the differences between the target values and the actual values.\footnote{Training normally ends after a given number of iterations or once a predefined error threshold has been met. The latter would make use of the computed error, based on the list \texttt{Delta}, but the corresponding code is omitted as it is not conceptually interesting.} The function \texttt{forward} computes the dot product of the input values and the weights of the outgoing edges. It is called twice by the function \texttt{ann} because there are two transitions between layers, first from the input layer to the hidden layer, and afterwards from the hidden layer to the output layer. Computing the dot product maps nicely to a functional programming style, as the required computation is the element-wise multiplication of two lists, followed by the summation of the results of that computation. The backpropagation pass starts with computing the output error, zipped with a \emph{squashing factor}. In our case, the activation function used for that purpose is a standard sigmoid function, the logistic function $f(x) = \frac{1}{1 + e^{-x} }$. The derivative of the logistic function is $f'(x) = f(x) (1 - f(x))$. This makes it possible to efficiently compute gradients as we can use the activations of the hidden layer for computing the total error in the output layer. Computationally, the operations involved, multiplication and subtraction, are less costly than re-evaluating the activation function, which is an exponential function.

\begin{figure}
\begin{lstlisting}[language=Erlang, caption=ANN helper functions, label=annHelper]
forward(_    , []      , Acc) -> lists:reverse(Acc);
forward(Input, [W | Ws], Acc) ->
  Val = lists:sum(lists:zipwith(fun(X, Y) -> X * Y end, Input, W)),
  forward(Input, Ws, [Val | Acc]).

output_error(Vals, Target) ->
  lists:zipwith(fun(X, Y) -> X * (1.0 - X) * (X - Y) end, Vals, Target).
  
backpropagate(_ , []    , []      , Acc) -> lists:reverse(Acc);
backpropagate(In, [E|Es], [Ws|Wss], Acc) ->
  A = lists:zipwith(fun(W, I) -> W - (E * I) end, Ws, In),
  backpropagate(In, Es, Wss, [A|Acc]).
  
errors_hidden([]    , _         , _      , Acc) -> lists:reverse(Acc);
errors_hidden([H|Hs], Output_Err, Weights, Acc) ->
  Outgoing = [ hd(X) || X <- Weights ],
  % Remaining weights for next iteration:
  Rest = [ tl(X) || X <- Weights ],
  % Error of current hidden layer neuron:
  TMP  = lists:zipwith(fun(X, E) -> E * X end, Outgoing, Output_Err),
  A    = lists:sum(TMP) * H * (1.0 - H),
  errors_hidden(Hs, Output_Err, Rest, [A|Acc]).

wrap_ann([]    , Weights, []    , Errors) ->
  {lists:reverse(Errors), Weights};
wrap_ann([I|Is], Weights, [T|Ts], Errors) ->
  { Error, Weights_ } = ann(I, Weights, T),
  wrap_ann(Is, Weights_, Ts, [Error | Errors]).
\end{lstlisting}
\end{figure}

The function \texttt{backpropagate} performs backpropagation, which computes the adjusted weights of the edges connecting the output layer to the hidden layer, and the adjusted weights of the edges connecting the hidden layer to the input layer. The new weights are computed by adding the product of the error and the input to each weight. The computation of the errors of the hidden layer is slightly trickier, due to using the list data structure. The weights assigned to the edges connecting the hidden layer with the output layer are specified as a list of lists in which each list contains the incoming weights of one of the output neurons. In the backpropagation pass, however, we need to traverse the ANN the opposite way, so the edges connecting the hidden layer to the output layer need a representation that considers all edges that point from the output layer to each node in the hidden layer. This is achieved by recursively taking the heads of the list of lists of the weights before performing the error calculation. Lastly, performing training on the entire input, so-called batch training, can be elegantly expressed in a functional style, shown by the function \texttt{wrap\_ann}. Its arguments are, in order, the list of inputs that constitute the training set, the weights, the target values associated with the input data, and an accumulator \texttt{Errors} that collects the output error for each element of the input set. The weights are continually updated so that every invocation of the function \texttt{ann} uses the weights of the preceding invocation.

\subsection{The Combined Framework}
\label{Sol3}
The parts introduced earlier can be combined to build a distributed system for Federated Learning. It boils down to using the skeleton introduced in Section~\ref{Sol1} and adding code for an artificial network to the client process, similar to what we have shown in Section~\ref{Sol2}, as well as further program logic. What has not been covered is, for instance, code for input/output handling. Our assumption is that each client process operates on data that is only locally available. The client process needs to be adjusted correspondingly, so that the available data is processed for batch-training with the ANN. Likewise, the server process needs to process the incoming models from the clients to update the centrally maintained global model, for instance via averaging.

While the description of our implementation is exclusively in Erlang, an alternative approach consists of a C implementation of the ANN. From a user perspective, there is no difference with regards to the output. Of course, internally the client trains an ANN in C instead of Erlang. However, in order to fairly compare how well an implementation solely in Erlang compares against one in which the computationally heavy lifting is performed by C, it is necessary to take the respective idiosyncrasies of two common approaches to interoperability with C into account. The older and more established way of calling C from Erlang is via so-called Natively Implemented Functions (NIFs), which are an improvement over using ports to communicate with C. A more recent addition to Erlang are C nodes, which have the advantage that they can be interfaced with the same way as regular Erlang nodes. Overall, for the purposes of simulating the framework, concurrent execution is adequate. However, distributed execution, in which messages are sent back and forth between nodes, more closely relates to real-world use cases (cf. Section \ref{prob_3}).

\subsection{Evaluation}
\label{Sol4}

\subsubsection{Setup.}
\label{Exp1}
We created four versions of our combined Federated Learning framework: (1) a concurrent implementation, fully in Erlang, as well as (2) one in which the clients are implemented in C as NIFs. Furthermore, we implemented (3) a distributed version fully in Erlang as well as (4) a variant of it in which the clients are C nodes. By default, all Erlang nodes in a distributed system are fully connected. As this is neither practical nor desirable for our use case, this behavior was disabled with the flag \texttt{-hidden}. Erlang source code has also been compiled to native code, which has been made possible due to the HiPE project~\cite{johansson2000high, sagonas2003all}.

The motivation behind benchmarking a distributed system, as opposed to the simpler case of a  concurrent system, is that this mirrors the real-world scenario of performing distributed data analytics tasks on a network with many edge devices and a central server. On the other hand, a concurrent system is more straightforward to design and execute. In both the distributed and the concurrent use case, we did not create our own implementation of a neural network in C. Instead, we chose Nissen's widely used Fast Artificial Neural Network (FANN) library~\cite{nissen2003implementation} with the option \texttt{FANN\_TRAIN\_BATCH}, which uses gradient descent with backpropagation. This corresponds to our Erlang code. Our ANN implementation in Erlang mirrors the chosen architecture of the ANN in FANN, i.e. there are two input nodes, three hidden nodes, and two output nodes. Furthermore, there is one bias node each, connecting to the hidden and the output layer, respectively. Error computation is done via computing the mean squared error (MSE) in both implementations. The Erlang code does not use a learning rate $\eta$, which implies that $\eta = 1$. In FANN, $\eta$ was explicitly set to~1 in order to override the default value of 0.7. Both implementations use the sigmoid activation function~(cf. Section~\ref{Sol2}). The corresponding setting in FANN is \texttt{FANN\_SIGMOID}.

\subsubsection{Hardware and Software.}
We used a PC with an Intel Core i7-7700K CPU clocked at 4.2 GHz. This is a quad-core CPU that supports hyper-threading with 8 threads. Our code was executed in Ubuntu Linux 16.04 LTS on a VirtualBox 5.1.22 virtual machine hosted by Windows 10 Pro (build 1703). The total amount of RAM available on the host machine was 32 GB, of which 12 GB were dedicated to VirtualBox. We used Erlang/OTP 20.2.2 and, for C, GCC 5.4.0.

\subsubsection{Experiment.}
For benchmarking novel machine learning methods, standard data sets are often used. These include the Iris data set~\cite{fisher1936iris}, which contains observational data of the petal length of various iris species. A more ambitious data set is the MNIST handwritten digits database~\cite{lecun2010mnist}. However, our goal was to directly compare the performance of two pairs of systems, so it seemed more appropriate to generate an artificial data set. Our data set is based on the mathematical function $f(x,y) = (\sqrt{xy}, \sqrt[4]{xy})$, where $\{x,y \in\mathbb{R} \mid x, y \in[0, 1)\}$. The ANN consists of two input nodes, three hidden nodes, and two output nodes. The training data consists of tuples $(x,y)$. Each client randomly generated 250 such tuples prior to each round of training. As the relationship between input and output is known, it is trivial to generate an arbitrary amount of data. We performed five 500-second test runs with each of the four combined frameworks, recording time, number of executed iterations of the ANN on each client node, and total error.  We used 10 client processes and hard-coded the initial weights for the sake of easy reproducibility. An alternative approach would have been to create the initial weights with the same random seed. In practice, the difference is insignificant.

The number of clients may seem small. However, the client part of our system will eventually be executed on separate hardware, such as the aforementioned on-board units in connected vehicles, which would each represent one single client node. Consequently, the focus is not on how the performance of \texttt{ffl-erl} scales when adding increasing numbers of client nodes to one machine. Given the recent interest in deep learning, one may also question the choice of using a shallow ANN. Shallow ANNs are still viable, however. In our case, the specified function is approximated successfully. On a related note, an interesting recent example of using a shallow ANN for computationally challenging work was presented by Cuccu et.\ al~\cite{cuccu2018playing}. They show that a shallow ANN can, in some tasks, compete with deep neural networks.

\subsubsection{Results.}
\label{Exp2}
Experimental results are shown in Fig.~\ref{fig:concurrent} below. The $x$-axis indicates the running time in seconds, while the $y$-axis shows the number of epochs, i.e.\ the number of iterations of the server-side ANN. Clients perform batch training on 250 data points per epoch. Each training pass consists of a constant amount of work, so the expectation was that the final result would be linear. The plotted data is the average of five test runs, which yielded virtually identical results. 

In the concurrent use case, the Erlang-only implementation, compiled to the BEAM virtual machine, executes $\sim$192,000 epochs in 500 seconds. This value increases to $\sim$286,000 epochs when compiling Erlang to native code. In comparison, the result with NIFs is $\sim$386,000 epochs. NIFs cannot be used together with natively compiled Erlang code, which is why a corresponding plot is missing. The performance difference between using NIFs and Erlang code compiled to BEAM amounts to a constant factor of 2.01. With native execution, the speedup compared to execution on the BEAM virtual machine amounts to 49.0\%. Comparing that performance to NIFs, the resulting difference shrinks to a factor of 1.35. 

An Erlang-only distributed implementation running on the BEAM virtual machine is able to compute $\sim$128,000 epochs in 500 seconds, which increases to $\sim$250,000 epochs (+95.3\%) with native code. On the other hand, with C nodes, the resulting performance is $\sim$522,000 epochs on the BEAM virtual machine as opposed to $\sim$643,000 epochs~(+23.4\%) per client when compiling Erlang to native code. The performance difference between a pure Erlang implementation and one that uses C nodes amounts to a constant factor of 4.1 on the BEAM virtual machine, which shrinks to 2.6 with HiPE.

\begin{figure}[htbp]
\hspace{-0.15cm}
    \begin{subfigure}[b]{0.4\textwidth}
        \includegraphics[trim={0.8cm 0.4cm 0.3cm 0.3cm},clip, scale=0.40]{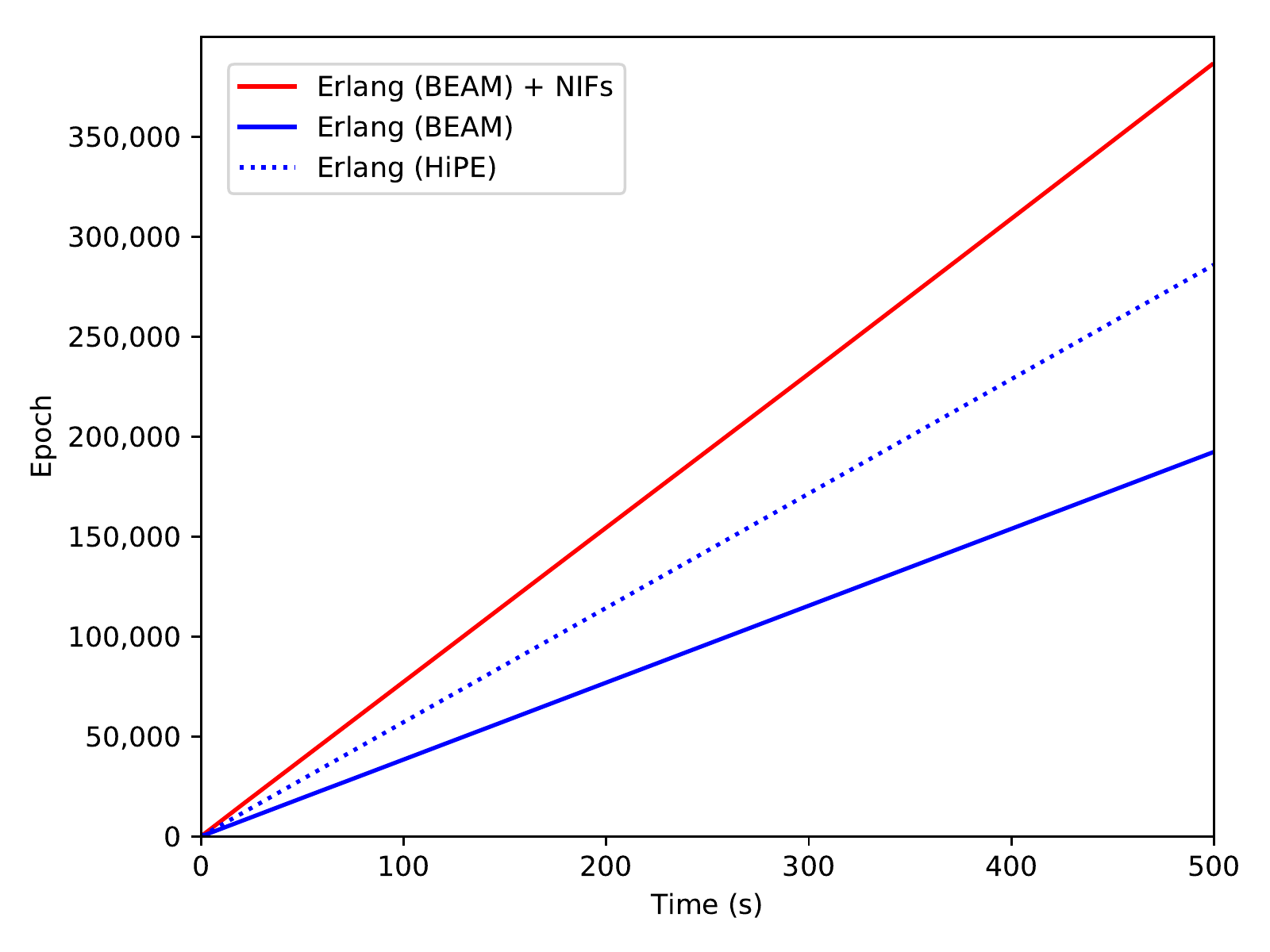}
        \caption{Concurrent execution}
        \label{rfidtest_xaxis}
    \end{subfigure}
    \hspace{11mm}
    \begin{subfigure}[b]{0.4\textwidth}
        \includegraphics[trim={0.8cm 0.4cm 0.3cm 0.3cm},clip, scale=0.40]{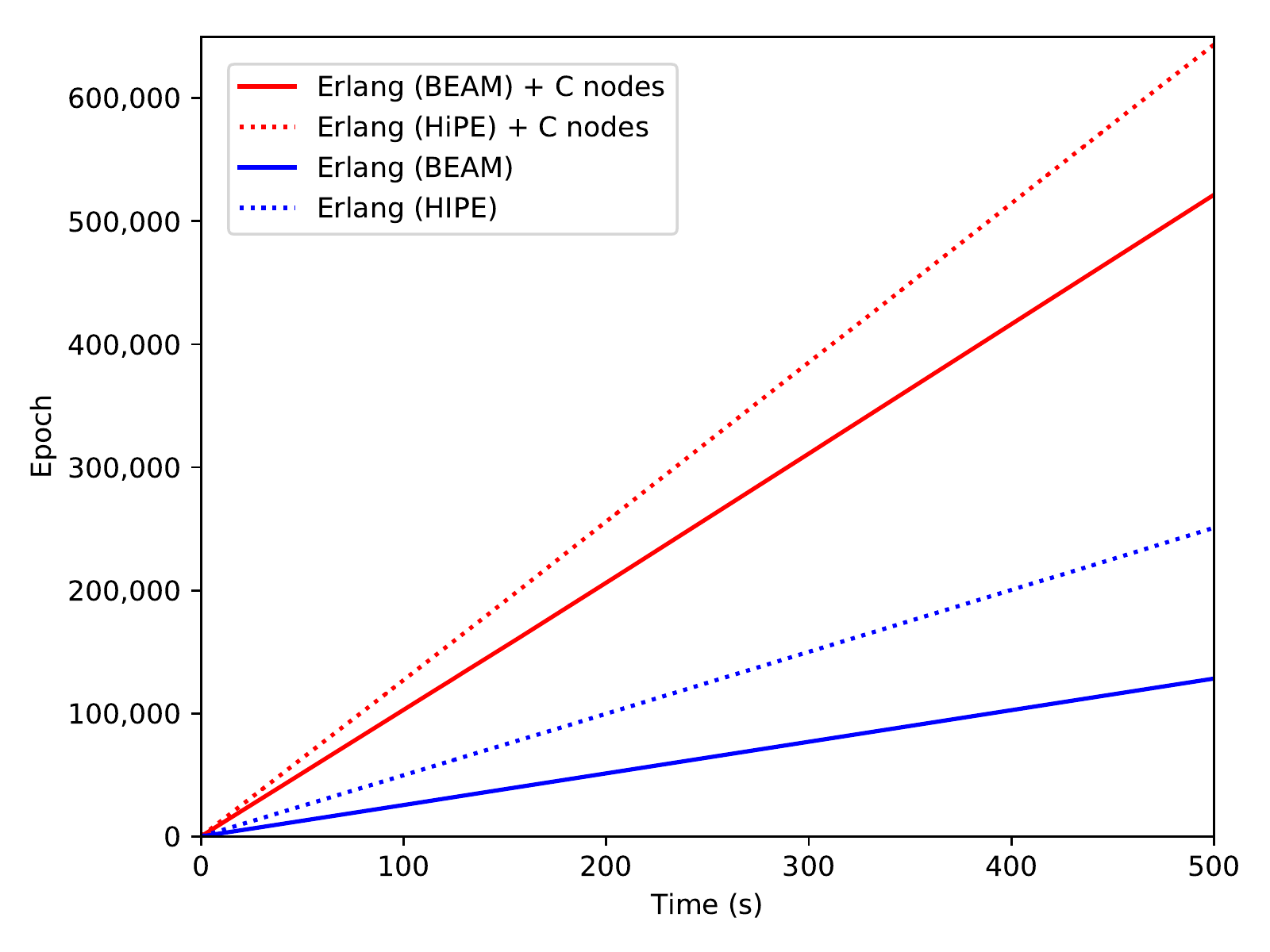}
        \caption{Distributed execution}
        \label{rfidtest_yaxis}
    \end{subfigure}
    \caption[]{In (a), Erlang (HiPE) reaches 74.1\% of the performance of Erlang code that uses NIFs. In (b), Erlang (BEAM) reaches 24.5\% of the performance of Erlang with C nodes; Erlang (HiPE) reaches 38.9\% of the corresponding performance.}
    \label{fig:concurrent}
\end{figure}

\subsubsection{Discussion.}
\label{Exp3}
It is perhaps surprising that an implementation that relies on Erlang for numerical computations is fairly competitive with C, with the observed difference amounting to a modest constant factor. In particular, the performance of natively compiled Erlang code is commendable. In the concurrent use case in particular, HiPE performs remarkably well. These are significant results for a number of reasons. From the perspective of programmer productivity, the relative conciseness of Erlang code, compared to C, is worth pointing out. For instance, the line count of our C code that merely interfaces with the FANN library slightly exceeds the line count of the Erlang implementation of our entire ANN. Writing the former was more time-consuming than the latter. That being said, the tool Nifty~\cite{loscher2016nifty}, which automates the generation of NIF libraries based on C header files, may have simplified this task. However, as we wanted to limit external dependencies, this was not a viable option.

The main argument for using C is its high performance. Yet, a downside is that it is a low-level programming language. In particular, manual memory allocation and garbage collection are an abundant source of programming errors. In terms of programmer productivity, C therefore does not compare favorably with Erlang. As there are use cases where performance is not of the topmost priority, Erlang may be a viable alternative as it leads to a much shorter turnaround time between design, implementation, and execution.

The performance comparison between Erlang and C is arguably lopsided, due to using the open-source C library FANN. It originally appeared in 2003 and has been actively maintained for over a decade, even though development activity seems to have slowed down recently. On the other hand, we developed our Erlang implementation of an ANN relatively quickly and without the benefit of extensively using it in real-world situations. Because FANN has been much more optimized than our code, the true performance difference between the competing programming languages may be less than what our numbers indicate.

C nodes work very well as they can essentially be addressed like Erlang nodes. NIFs, on the other hand, have serious drawbacks.\footnote{Refer to the section on Natively Implemented Functions (NIFs) in the official Erlang documentation for further details: \url{http://erlang.org/doc/man/erl\_nif.html} (accessed on June 28, 2018).} They are executed as native extensions of the Erlang VM. Thus, a NIF that crashes will crash the Erlang VM. Furthermore, NIFs can cause state inefficiencies, which may lead to crashes or unexpected behaviors. Lastly, there is the issue of \emph{lengthy work}: a NIF that takes too long to return may negatively affect the responsiveness of the Erlang VM. In the Erlang version we were using, a well-behaving NIF has to return within one millisecond. In exploratory benchmarking with data sets not much larger than the one we eventually used, we measured calls to NIFs that took longer than that. Consequently, we think it is too risky to use NIFs in a more taxing environment.

\section{Related Work}
\label{related}
There has been some preceding work in academia related to using functional programming languages for tackling machine learning tasks. About a decade ago, Allison explored using Haskell for defining various machine learning and statistical learning models~\cite{allison2005models}. Yet, that work was of a theoretical nature. Going back even further, Yu and Clack presented a system for polymorphic genetic programming in Haskell~\cite{yu1998polygp}. Likewise, this was from a theoretical perspective. More recently, Sher~\cite{sher2012handbook} did extensive work on modeling evolutionary computations. A central part of his contribution is an ANN implemented in Erlang. However, his fairly complex system could only have been used as the starting point of our work with substantial modifications. One key difference is that individual nodes of the ANN are modeled as independent processes, and so are sensors and actuators. A related ANN implementation in Erlang is \texttt{yanni},\footnote{The corresponding code repository is located at \url{https://bitbucket.org/nato/yanni} (accessed on August 6, 2018).} which follows Sher's approach of using message passing, albeit only between layers.

\section{Future Work}
\label{future}

The \texttt{ffl-erl} project has influenced ongoing work in our research lab on a real-world system for distributed data analytics for the automotive industry~\cite{ulm2019oodida}. In that system, Erlang is used for distributing assignments to clients, which operate on local data. Those clients can execute code written in an arbitrary programming language. Federated Learning is one of its use cases.

\subsubsection*{Acknowledgements.} 
Our research was financially supported by the project On-board/Off-board Distributed Data Analytics (OODIDA) in the funding program FFI: Strategic Vehicle Research and Innovation (DNR 2016-04260), which is administered by VINNOVA, the Swedish Government Agency for Innovation Systems. It was carried out in the Fraunhofer Cluster of Excellence "Cognitive Internet Technologies." Adrian Nilsson and Simon Smith assisted with the implementation. Melinda T\'oth pointed us to Sher's work. We also thank our anonymous reviewers for their helpful feedback.
%
%

\bibliographystyle{splncs04}


\appendix
\section{Mathematical Derivation of  Federated Stochastic Gradient Descent}
\label{appendix_A}
In Section~\ref{sub:fsgd} we briefly describe Federated Stochastic Gradient Descent. In the current section, we present the complete derivation. As a reminder, we stated that in Stochastic Gradient Descent, weights are updated this way:

\begin{equation} \label{eq:X}
w := w - \frac{\eta}{n} \displaystyle\sum_{i=1}^{n} \nabla F_i(w).
\end{equation}

Furthermore, we started with the following equation, which is the objective function we would like to minimize:

\begin{equation}
F(w) = \frac{1}{n} \displaystyle\sum_{j=1}^{k} |P_j| F^{j}(w).
\end{equation}

The gradient of $F^{j}$ is expressed in the following formula:

\begin{equation}
\nabla F^{j}(w) = \frac{1}{|P_j|} \displaystyle\sum_{i \in P_{j}}^{} \nabla F_i(w), j = 1,\dotsc,k.
\end{equation}

To continue from here, each client updates the weights of the machine learning model the following way:

\begin{equation}
w_j = w - \frac{\eta}{|P_j|} 
\displaystyle\sum_{i \in P_{j}}^{}
\nabla F_i(w).
\end{equation}

On the server, the weights of the global model are updated. The original equation can be reformulated in a few steps:

\begin{align}
w :=& \frac{1}{n}\left(\displaystyle\sum_{j=1}^{k} w_j |P_j|\right) \\
   =& \frac{1}{n}\displaystyle\sum_{j=1}^{k} \left(w - \frac{\eta}{|P_j|} 
\displaystyle\sum_{i \in P_{j}}^{} \nabla F_i(w)\right) |P_j|\\
=& \frac{1}{n}\displaystyle\sum_{j=1}^{k} |P_j|w -
  \frac{1}{n}\eta
  \displaystyle\sum_{j=1}^{k}
  \displaystyle\sum_{i \in P_{j}}^{}
  \nabla F_i(w)
   \label{eq:from}\\
=& w - \frac{\eta}{n}
    \displaystyle\sum_{i=1}^{n} \nabla F_i (w).
     \label{eq:to}
\end{align}

The reformulation in the last line is equivalent to Equation~\ref{eq:X} above. In case the transformation between Eq.~\ref{eq:from} and Eq.~\ref{eq:to} is unclear, consider that the summand simplifies to

\begin{equation}
\frac{1}{n}\displaystyle\sum_{j=1}^{k} |P_j|w
= \frac{1}{n} n w = w.
\end{equation}

The second summand in Eq.~\ref{eq:from} can be simplified as follows:

\begin{equation}
  \frac{1}{n}\eta
  \displaystyle\sum_{j=1}^{k}
  \displaystyle\sum_{i \in P_{j}}^{}
  \nabla F_i(w)  
  = \frac{1}{n}\eta
  \displaystyle\sum_{i = 1}^{n}
  \nabla F_i(w)
 = \frac{\eta}{n}
  \displaystyle\sum_{i = 1}^{n}
  \nabla F_i(w).
\end{equation}

\end{document}